\documentclass[twocolumn,aps,showpacs,prl,epsf,graphics,psfig]{revtex4}
\usepackage{graphicx}

\begin{document}
\title{Piezomagnetic Quantum Dots}

\author{Ramin M. Abolfath$^1$}
\author{Andre Petukhov$^2$}
\author{Igor \v{Z}uti\'c$^1$}

\affiliation{
$^1$ Department of Physics, State University of New York at Buffalo,
Buffalo, New York 14260, USA \\
$^2$  Department of Physics, South Dakota School
of Mines and Technology, Rapid City, SD 57701-3995
}

%\date{\today}

\begin{abstract}
We study the influence of deformations on magnetic ordering in quantum 
dots doped with magnetic impurities. 
The reduction of symmetry and the associated deformation from 
circular to elliptical quantum confinement lead to the formation 
of piezomagnetic quantum dots. The strength of elliptical deformation
can be controlled by the gate voltage 
to change the magnitude of
magnetization, at a fixed number of carriers and in the absence of
applied magnetic field. 
We reveal a reentrant magnetic ordering with the increase of elliptical 
deformation and suggest that the piezomagnetic quantum dots can be used
as nanoscale magnetic switches.
\end{abstract}
\maketitle
Quantum dots (QDs) can be viewed as artificial atoms which allow for
a versatile control of the number of carriers, their spin, and the effects of
quantum confinement~\cite{Reimann2002:RMP,Jacak1998:Book,Maekawa:2006}. 
The motivation to 
magnetically dope 
semiconductor QDs comes from the possibility for an enhanced
control of the magnetic ordering as compared to their bulk-like 
counterparts~\cite{Ohno2000:N,Oiwa2002:PRL,%
Jungwirth2006:RMP,Zutic2004:RMP}
as well as to directly study the influence of quantum confinement
and strong Coulomb interactions on magnetism. There is an
encouraging experimental progress in Mn-doped II-VI and 
III-V QDs, from the controlled inclusion of a single Mn impurity
to the onset of magnetization at temperatures
substantially higher than in similar bulk 
materials~\cite{Mackowski2004:APL,Besombes2004:PRL,Leger2006:PRL,%
Gould2006:PRL,Chakrabatri2005:NL,Archer2007:NL}.
However, most of the theoretical studies have been limited to 
a small number of carriers 
($N$) and Mn-impurities ($N_m$) and the effects of the symmetry of quantum 
confinement and electron-electron (e-e) Coulomb interaction on the magnetic 
phase diagram, remain to be understood~\cite{Fernandez-Rossier2004:PRL,%
Govorov2005:PRB,Qu2005:PRL,Abolfath2007:PRL,Zhang2006:P}.
In particular, it would be useful to know what 
type of external influences can effectively produce magnetic ordering
of carrier spins and magnetic impurities and how such ordering would modify
the electrical, optical, and transport properties of QDs. 

In the absence of magnetic doping, circular
QDs with rotational symmetry of the lateral confining potential 
have degenerate energy levels and show 
a pronounced shell structure, similar to atoms and nuclei.  
The interplay of the Pauli principle and the reduction 
of e-e Coulomb exchange energy near degenerate levels in the single-particle 
spectrum, according to Hund's first rule, favors the parallel spin alignment 
of electrons in open shells~\cite{Jacak1998:Book,Reimann2002:RMP}.
Breaking the circular symmetry by deforming the lateral confinement
removes the single-particle level degeneracies and 
leads to transitions in the spin of electrons, referred to as 
piezomagnetism~\cite{Reimann2002:RMP,Austing1999:PRB}. 
In vertical QDs, elliptical deformation can be 
made by 
micron-sized device 
mesas incorporating dots, where the lateral electrostatic confinement 
originates from side-wall depletion. Both the symmetry of the  
confinement and the effective QD size can be controlled by the action 
of a Schottky gate wrapped around the mesa in the vicinity of the 
QD~\cite{Austing1999:PRB,Matagne2002:PRB,Tarucha1996:PRL}.
In lateral QDs, the confinement potential is formed
electrostatically by several gate voltages, altering the shape and the 
deformation of the confining potential while keeping 
$N$ constant~\cite{Kyriakidis2002:PRB,Zumbuhl2004:PRL}.

We focus here on the effects of elliptical deformations on magnetic 
ordering in (II,Mn)VI QDs. 
In particular, we are interested to explore the onset of a spin-polarized
state with finite Mn-magnetization, in the absence of applied
magnetic field, to which we refer as ``ferromagnetic'' (FM) 
state~\cite{Fernandez-Rossier2004:PRL,Govorov2005:PRB,Qu2005:PRL}.
Since Mn is isoelectronic with group-II elements, the number
of carriers can be changed by chemical doping or by electrostatic gates.
We represent magnetic QD in zero magnetic field by the
Hamiltonian $H=H_e+H_{ex}+H_m$, describing contributions of
interacting electrons, electron-Mn (e-Mn) exchange, and
direct Mn-Mn antiferromagnetic (AFM) coupling, respectively.
The interacting electron contribution is given by 
\begin{eqnarray}
H_e = \sum_{i=1}^N 
[-\frac{1}{2m^*}\nabla^2_i 
+ U_{QD}({\bf r}_i)]  
+ \frac{e^2}{2\epsilon}\sum_{i\neq j} 
\frac{1}{|{\bf r}_i - {\bf r}_j|},
\label{eq1}
\end{eqnarray}
where we set $\hbar \equiv 1$, 
$m^*$ is the electron effective mass, 
${\bf r} \equiv (\vec{\rho},z)$,
and $U_{QD}({\bf r})= V_{QD}(\vec{\rho})+V^1_{QD}(z)$
is the three-dimensional (3D) QD confining potential,
where $V^1_{QD}(z) = m^*\Omega^2 z^2 / 2$ is the
1D parabolic potential with
the characteristic subband energy $\Omega$, and 
\begin{eqnarray}
V_{QD}(\vec{\rho})\equiv V_{QD}(x,y) 
=(1/2)m^*\omega_0^2 \left(x^2 \delta + y^2 \delta^{-1}\right),
\label{eq2}
\end{eqnarray}
is the 2D anisotropic parabolic potential~\cite{Reimann2002:RMP} 
which describes the lateral confinement with frequencies 
$\omega_x=\omega_0\sqrt{\delta}$, $\omega_y=\omega_0/\sqrt{\delta}$, 
where $\delta \equiv \omega_x / \omega_y \neq 1$ is the strength of 
elliptical deformation. We impose $\omega_0^2=\omega_x \omega_y$
which conserves the area of QD with deformation.
The last term in Eq.~(\ref{eq1}) is the repulsive 
e-e Coulomb interaction screened by the dielectric constant $\epsilon$,
and $-e$ is electron charge.
The e-Mn exchange contribution is 
$H_{ex}=- J_{sd}\sum_{i,I}\vec{s}_i\cdot\vec{M}_I  
\delta({\bf r}_i - {\bf R}_I)$,
where $J_{sd}$ is the exchange coupling between electron spin 
$\vec{s}_i$ at ${\bf r}_i$
and impurity spin $\vec{M}_I$ at ${\bf R}_I$.
The Mn Hamiltonian is
$H_m=\sum_{I,I'}J^{AF}_{I,I'} \vec{M}_I \cdot \vec{M}_{I'}$, where 
$J^{AF}_{I,I'}$ is the direct Mn-Mn AFM coupling.
The $z$-component of $\vec{M}_I$ is
$M_z=-M, -M+1, \dots, M$, where we choose $\hat{z}$ as the quantization 
axis and $M=5/2$ for Mn.
We use a real space finite-temperature local spin density approximation
(LSDA) and the mean filed approximation for Mn spins to avoid 
computational complexity~\cite{Abolfath2007:PRL,complex} 
of exact diagonalization,
limited to only a very small number
of interacting electrons and magnetic impurities~\cite{Qu2005:PRL}.
An effective Hamiltonian describing electrons can be obtained 
by replacing the Mn spins, that are randomly distributed, with a classical
continuous field
$H^{\rm eff}_e = H_e - \sum_i J_{sd} n_m \frac{\sigma_i}{2} 
\langle M_z({\bf r}_i)\rangle$,
where $n_m$ is the averaged density of Mn, and $\sigma=\pm 1$ for 
spin up ($\uparrow$), and down ($\downarrow$).
Within the mean field approach~\cite{Abolfath2007:PRL},
the magnetization density can be expressed as
$\langle M_z({\bf r}_i)\rangle = M B_M(M b({\bf r}_i)/k_BT)$ where
$B_M(x)$ is the Brillouin function~\cite{Furdyna1988:JAP}, 
$k_B$ is the Boltzmann constant, $T$ the absolute temperature,
while 
$b({\bf r}_i)=- J^{AF}_{\rm eff} \langle M_z({\bf r}_i)\rangle
+ J_{sd} [n_\uparrow({\bf r}_i) - n_\downarrow({\bf r}_i)]/2$
is the effective field seen by the Mn. 
The first term in $b({\bf r}_i)$ describes the mean field
of the direct Mn-Mn AFM coupling~\cite{Fernandez-Rossier2004:PRL}
and $n_\sigma({\bf r}_i)$ is the spin-resolved electron density.

Following the decomposition of 3D confining potential, 
we expand the QD wave functions in terms of its 
planar $\psi_{i\sigma}(\vec{\rho})$ 
and subband wave function $\xi(z)$ (in typical disk-shaped 
QDs only the first subband is filled).
We project $H^{\rm eff}_e$ into a 2D Hamiltonian
by integrating out $\xi(z)$.
In LSDA we express the Kohn-Sham (KS) Hamiltonian as
\begin{eqnarray}
H_{KS} = -\frac{1}{2m^*} \nabla_\rho^2
+ V_{QD} + V_{H} + V^\sigma_{XC}
- \frac{\sigma}{2} h_{sd},
\label{Heff}
\end{eqnarray}
where the $V_H$ is the Hartree potential,
$V^\sigma_{XC}$ is Vosko-Wilk-Nusair spin dependent 
exchange-correlation potential~\cite{Dharma-wardana1995:Book}, and
the exchange spin splitting is given by
\begin{eqnarray}
h_{sd}(\vec{\rho},T) =  J_{em} \int dz |\xi(z)|^2 
B_M\left( \frac{M b(\vec{\rho}, z)}{k_BT}\right).
\label{hsd}
\end{eqnarray}
where $J_{em} = J_{sd} n_m M$ is the e-Mn exchange coupling.
We solve self-consistently the 
Kohn-Sham equations, 
$H_{KS} \psi_{i\sigma}(\vec{\rho}) = 
\epsilon_{i\sigma}\psi_{i\sigma}(\vec{\rho})$,
where $H_{KS}$ is given by Eq.~(\ref{Heff}) and $\epsilon_{i\sigma}$ are 
the KS energies. 
Our numerical results are illustrated for parameters 
based on (Cd,Mn)Te QDs with $J_{sd}=0.015$ eV nm$^3$, $m^*=0.106$, 
$\epsilon=10.6$~\cite{Qu2005:PRL}, $a^*_B=5.29$ nm, and $Ry^*=12.8$ meV 
are the effective Bohr radius and Rydberg energy.
We choose a QD with $\omega_0=25.6$ meV,
$n_m=0.1$ nm$^{-3}$, $J^{AF}_{\rm eff} = 0.02$ meV, and having 
the perpendicular ($z$) width 1 nm.
\begin{figure}
\begin{center}\vspace{0.2cm} 
\includegraphics[width=0.9\linewidth]{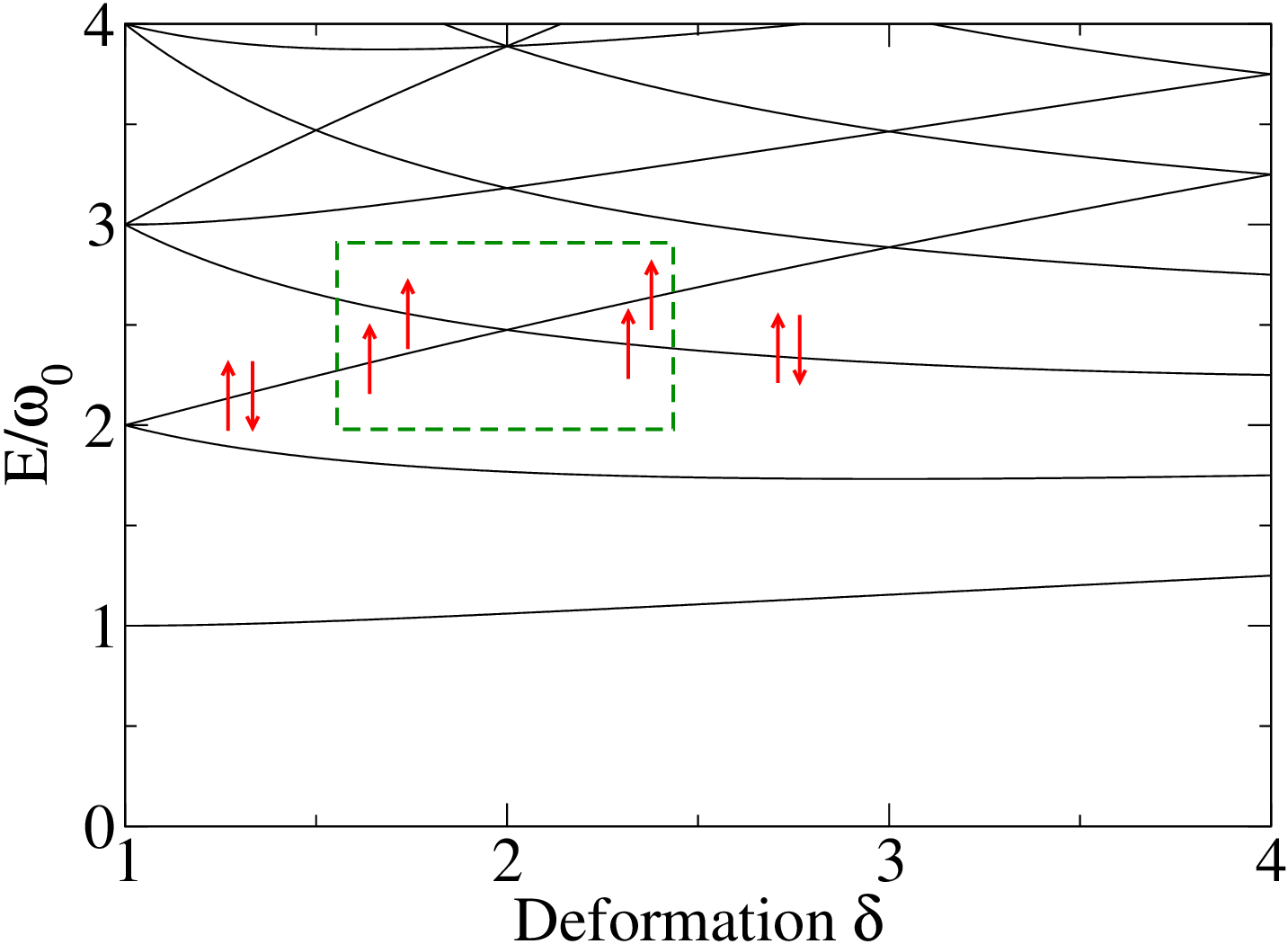}\vspace{0.9cm}
\includegraphics[width=0.9\linewidth]{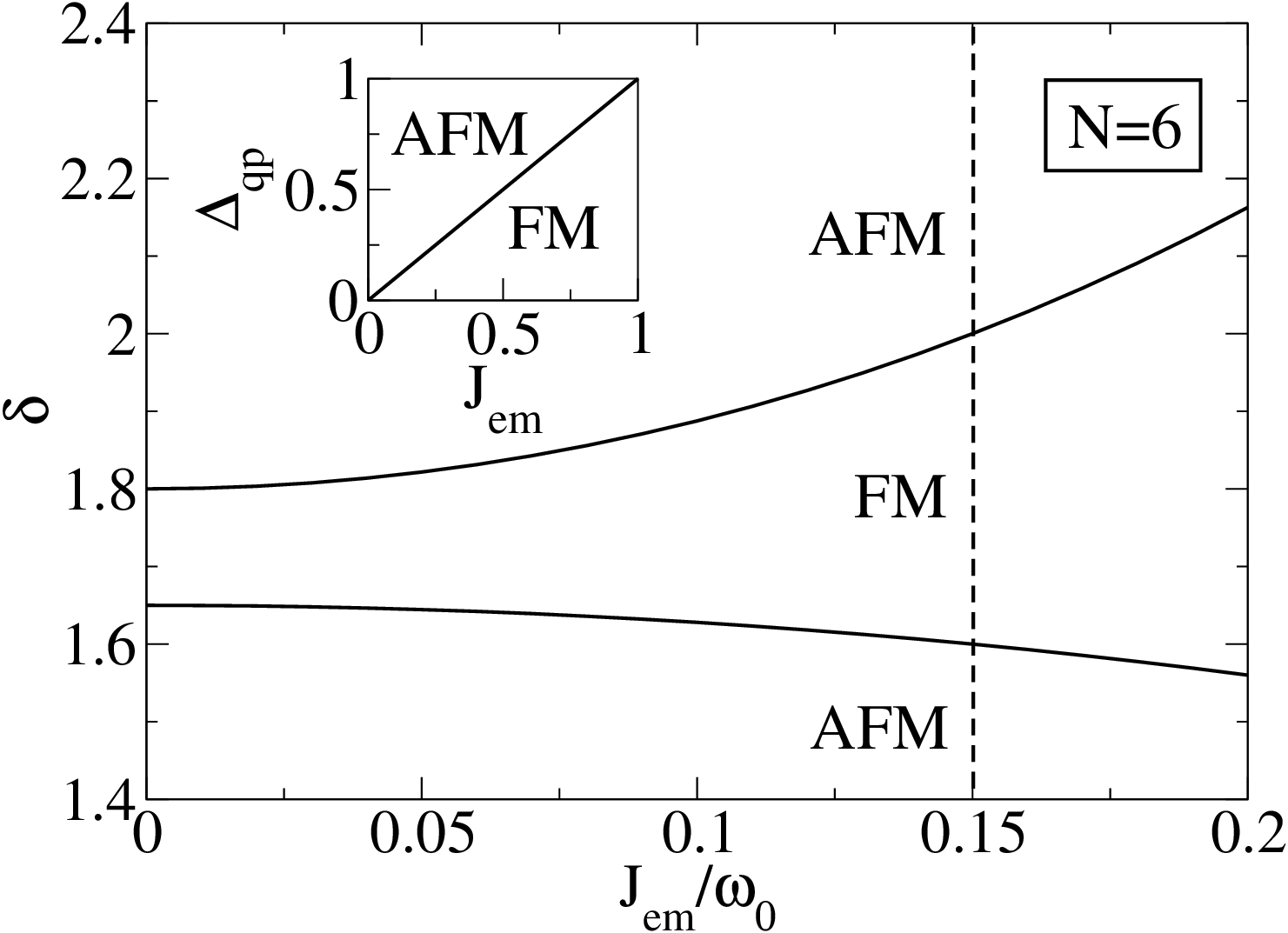}
\caption{
Effects of elliptical deformation.
Top: 
Single-particle levels of the anisotropic QD as a 
function of  deformation $\delta$. 
Degeneracies lead to the formation of closed shells with 
$N=2,6,12,20,..$,  in the isotropic (circular) case 
($\delta=1$), and the formation of subshells at $\delta=q/p$ for 
integer $q,p$. Broken square and the corresponding 
arrows schematically illustrate how the spin alignment for $N=6$ would be 
modified near the degeneracy point at $\delta=2, E/\omega_0=2.5$ 
in the presence of e-e Coulomb interaction and magnetic 
impurities. 
Bottom: The inset illustrates the magnetic phase diagram
as a function of quasi-particle gap $\Delta_{qp}$, 
and e-Mn exchange coupling $J_{em}$.  The line $\Delta_{qp}=J_{em}$ 
separates ferromagnetic (FM) and antiferromagnetic (AFM) ordering.
The magnetic phase diagram of QD with $N=6$ interacting
electrons at $T=0.5$ K is calculated as a function of 
$\delta$ and $J_{em}/\omega_0$.
The dashed line corresponds to $n_m=0.1$ nm$^{-3}$ and $\omega_0=25.6$ meV.
}
\label{ffig1}
\end{center}
\end{figure}

We first examine the effects of elliptical deformation,
described with the anisotropic 2D lateral confinement from Eq.~(\ref{eq2}),
on the single-particle levels. 
In the absence of e-e Coulomb interaction and magnetic 
impurities $J_{em}=n_m=0$,
the corresponding single-particle spectrum 
$E_{n_x,n_y} (\delta) = \omega_0 [(n_x + 1/2)\sqrt{\delta} 
+ (n_y + 1/2)/\sqrt{\delta}]$, where
$n_{x(y)} = 0, \pm 1, \pm 2, \cdots$, 
is shown as a function of deformation $\delta$ in Fig.~\ref{ffig1}, top.
Isotropic QD ($\delta=1$) have a pronounced shell structure
with $n_x + n_y + 1$-fold degeneracy in each shell, 
having $N=2,6,12,20,..$ electrons.
While, any deviation from the isotropic parabolic confinement 
breaks the geometrical 
symmetry and removes these degeneracies, additional accidental degeneracies,
can occur in deformed QD at $\delta=p/q$ for
integer $p,q$ and lead to the formation of subshells.
For example, at $\delta=2$  both $p$- and $d$-, 
as well as $d$- and $f$-levels cross, as shown in Fig.~\ref{ffig1}, top.
The $\delta=2$ shell structure is significantly modified from the 
isotropic case,
$N=2,4,8,12,...$, correspond to the closed shells and 
$N=6,10,15,...,$ to the half-filled shells.

For the study of magnetic ordering in (II,Mn)VI QDs, it is important
to understand how this single-particle picture is modified in the presence
of e-e Coulomb interaction and e-Mn exchange coupling $J_{em}$. 
The corresponding sequence of filling electrons and their spin 
alignment, through carrier-mediated magnetism, affects the 
magnetic ordering of Mn-impurities and
the possibility to externally control Mn-magnetization.  

A simple sketch
of anticipated modification near $\delta=2$ for the $N=6$ state,
shows that the parallel spin alignment of the valence electrons is favored 
for a wide range of deformation.  Additional analysis is addressed in 
Fig.~\ref{ffig1}, bottom. The inset shows that the occurrence of FM or AFM state
(having finite and vanishing magnetization, respectively), depends on the 
competition between the two characteristic energy scales: the quasi-particle 
energy gap $\Delta_{qp}$, and $J_{em}$~\cite{afm}. 
The single-particle gap should be distinguished from
$\Delta_{qp}$ which is the energy difference between highest occupied 
and lowest unoccupied KS levels and depends on $N$, 
the total spin of electrons, 
the strength of e-e Coulomb interaction, and the underlying symmetries 
of QD confining potential. 
For $J_{em}=0$, as shown in Fig.~\ref{ffig1} 
the FM state is stable only within a narrow range of $\delta$ which 
becomes vanishingly small 
for noninteracting electrons (at $\delta=2$, Fig.~\ref{ffig1}, top). 
The increase in $J_{em}$ extends the deformation range
over which the FM state is possible.

\begin{figure}
\begin{center}\vspace{0.3cm}
\includegraphics[width=0.9\linewidth]{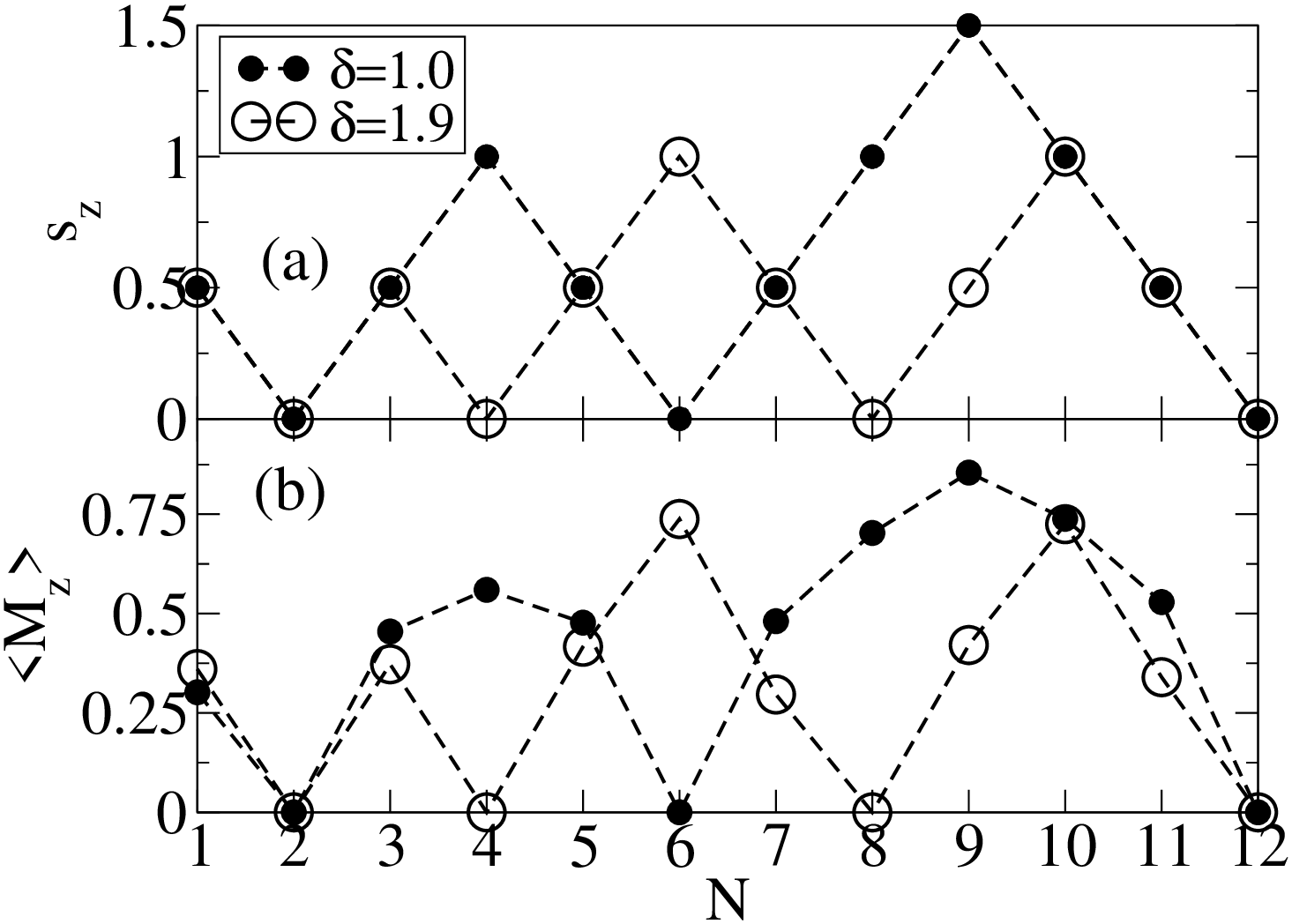}
\caption{
The $z$-component of the total spin of electrons $s_z$ (a) and 
averaged magnetization per unit area $\langle M_z \rangle$ (b) as
a function of number of electrons $N$ at $T=0.5$ K and the Mn-density
$n_m=0.1$ nm$^{-3}$. Isotropic ($\delta=1$, filled circles) and
elliptical QD ($\delta=1.9$, empty circles) show qualitatively different
dependence of $s_z$ and $\langle M_z \rangle$ with $N$.
}
\label{fig1}
\end{center}
\end{figure}

To further examine the effect of elliptical deformation on magnetic
ordering we introduce the
$z$-component of the total spin of electrons,
$s_z=(N_\uparrow - N_\downarrow)/2$, and
the spatially-averaged Mn-magnetization
per unit area $A$, $\langle M_z \rangle = (1/A) 
\int d^2\rho \langle M_z(\vec{\rho}) \rangle$,
obtained from the self-consistent solution of Kohn-Sham equations.
In Fig.~\ref{fig1} we show $s_z$ and
$\langle M_z \rangle$ as a function of $N$ for both isotropic QD 
with circular symmetry ($\delta=1$), and elliptical QD with 
$\delta=1.9$.
For interacting electrons and finite $J_{em}$ we observe that
carrier-mediated magnetism leads to coupled ordering of electron
and Mn spins. 
The oscillations of $s_z$  and $\langle M_z \rangle$ reveal that the shell
structure is strongly modified with deformation. 
The vanishing of $s_z$ and $\langle M_z \rangle$ correspond to the 
closed shells ($N=2,6,12,...$, at $\delta=1$ and $N=2,4,8,12,...$,
at $\delta=1.9$) while their local maxima correspond to the half-filled
shells ($N=1,4,9,...$, at $\delta=1$ and $N=1,3,6,10,...$,
at $\delta=1.9$). 

Focusing on the $N=6$ state, we see that the minima 
in $s_z$ and $\langle M_z \rangle$ at $\delta=1$ become
maxima at the deformation $\delta=1.9$. 
For an isotropic QD the $p$-levels are 
degenerate (Fig.~\ref{ffig1}) and $N=6$ corresponds to the AFM state
with $\Delta_{qp}>J_{em}$ and vanishing $\langle M_z \rangle$.
Elliptical deformation breaks the circular symmetry of
the lateral confinement and removes the $p$-level degeneracy. 
At $\delta=1.9$, $J_{em}>\Delta_{qp}$ stabilizes the FM state 
with finite $\langle M_z \rangle$ and $s_z=1$ implies that the
electron spin polarization is
$P=(N_\uparrow-N_\downarrow)/(N_\uparrow+N_\downarrow)=(4-2)/6$.
The $N=6$ state illustrates an intriguing possibility to turn the 
Mn-magnetization {\it on} and {\it off} without changing the number
of carriers, needed in the bulk-like magnetic semiconductors.
However, $N=6$ is not the unique state, a similar behavior 
near $\delta=1.5$ is expected also for the $N=12$ state, 
forming another closed shell
in the limit of isotropic confinement. 
Elliptical deformations of the lateral confinement
in Mn-doped QDs thus lead to the change between the AFM and FM state which
can be controlled by gate voltage. Such structures would behave as 
piezomagnetic QDs in which the control of deformation determines 
the magnetic ordering.

\begin{figure}
\begin{center}\vspace{1cm}
\includegraphics[width=0.9\linewidth]{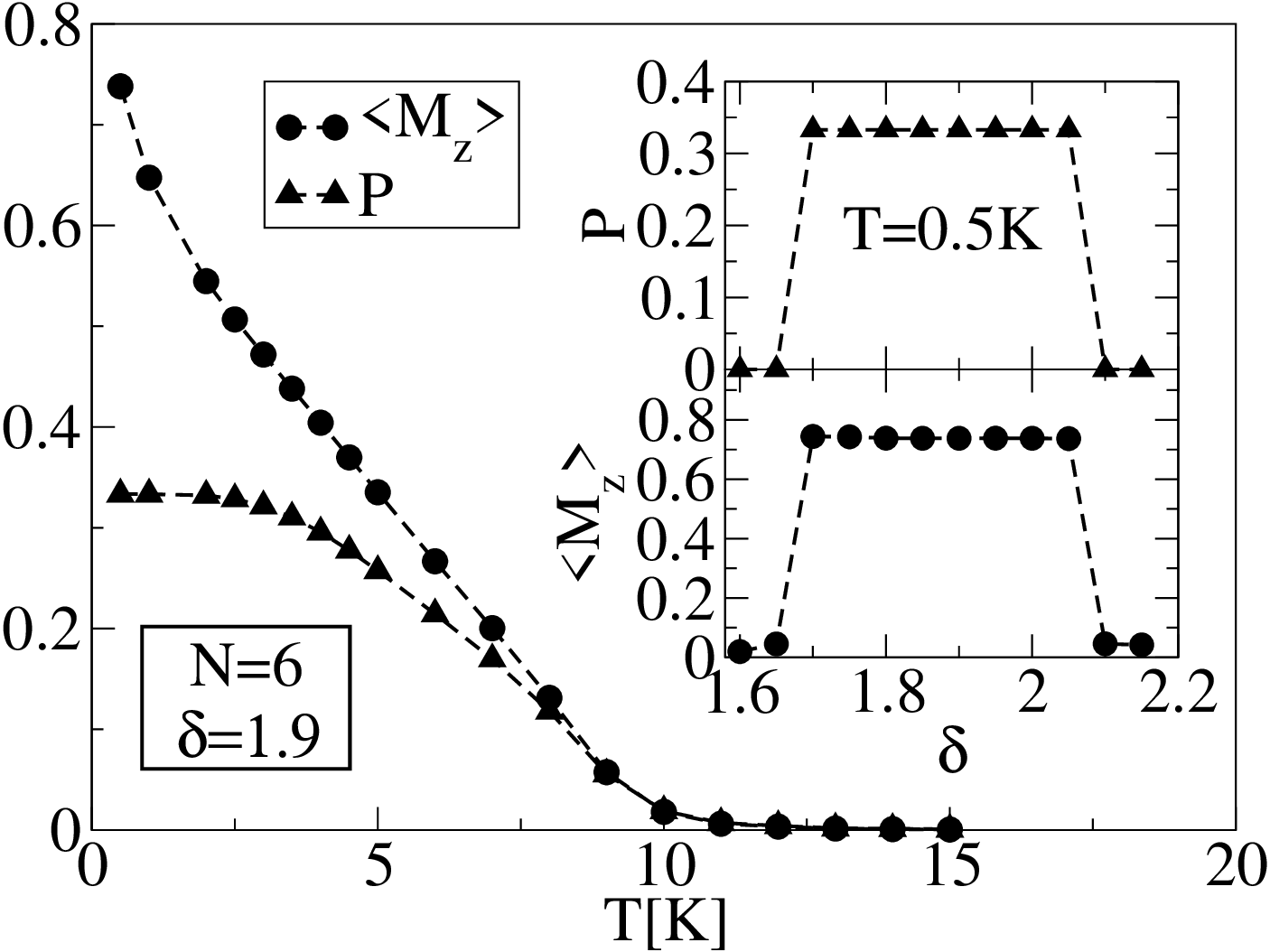}
\caption{
Temperature evolution of Mn-magnetization per unit area $\langle M_z \rangle$
and electron spin polarization $P$
in an elliptical QD with $\delta=1.9$ and $N=6$.
Inset: Evolution of 
$\langle M_z \rangle$ and $P$ with elliptical deformation at $T=0.5$ K
shows a nonmonotonic behavior with sharp rise
and fall (near $\delta=1.7$ and $\delta=1.9$, respectively) 
corresponding to the switching between the AFM and FM state which could
be controlled by gate voltage.
}
\label{fig4}
\end{center}
\end{figure}

To examine if the piezomagnetic QDs could 
function as nanoscale magnetization switches, we also consider 
the evolution of their magnetic ordering with $T$ and $\delta$, 
shown in Fig.~\ref{fig4} for $N=6$. While 
$\langle M_z \rangle$ and $P$ monotonically decay with $T$, 
the inset reveals a nonmonotonic behavior of 
$\langle M_z \rangle$ and $P$ with $\delta$. The reentrant magnetic
ordering with the increase of elliptical deformation 
corresponds to the AFM-FM-AFM transitions, anticipated from the sketch
in Fig.~\ref{ffig1}, top. Since the change between vanishing and finite 
$\langle M_z \rangle$ occurs in a narrow range of $\delta$ (near $1.7$
or $2.1$, respectively) we expect that a moderate change in the gate voltage
would suffice to implement such a magnetization switch~\cite{private}.
With the increase of $\delta$ we can also anticipate FM-AFM-FM transitions.
In the limit of $\delta=1$, a QD with $N=10$ 
forms the FM state with finite $\langle M_z \rangle$ and $s_z=1$ 
but with an increased deformation and opening up of $\Delta_{qp}$,   
near $\delta=1.5$, the AFM state with $\langle M_z \rangle=s_z=0$ 
is favored. With the further increase in $\delta$, 
we find switching back to the FM state with $s_z=1$, near $\delta=2$,
also expected
from a single-particle picture (Fig.~\ref{ffig1}, top) 
and the crossing of $d$- and $f$-levels at $\delta=2$.

Another potential application of piezomagnetic QDs could be their 
voltage-controlled $P$ to enable an efficient and tunable spin injection
in semiconductor nanostructures~\cite{Zutic2004:RMP}. 
While for the $N=6$ state the total 
electron spin polarization is limited to $P=(4-2)/6=1/3$, the 
transfer of only valence electrons could lead to the
injection of completely spin-polarized carriers, in the absence
of applied magnetic field. Analogously, such QDs could also
be used as voltage-controlled spin filters~\cite{Efros2001:PRL}. 
	
In conclusion, we have investigated the effects of elliptical deformation, 
electron-electron Coulomb interaction, and the number of carriers
on the magnetic ordering in Mn-doped QDs. 
We reveal additional possibilities to control magnetism
in semiconductor nanostructures which would not be feasible in 
their bulk counterparts.
The two  previous experimental results: 
(1) the feasibility of large elliptical deformations
without the Mn-doping~\cite{Austing1999:PRB} and (2) 
controlled Mn-doping in 
QDs~\cite{Mackowski2004:APL,Besombes2004:PRL,Leger2006:PRL,%
Gould2006:PRL,Chakrabatri2005:NL,Archer2007:NL},
would need to be combined in order to realize the
predicted behavior of piezomagnetic QDs. While  
our analysis has focused on the electrons and II-VI materials,
much larger hole-Mn exchange 
coupling~\cite{Furdyna1988:JAP,Bhattacharjee1997:PRB}
in either II-VI or III-V QDs 
could extend the suggested control and  manipulation of magnetic ordering to 
significantly higher temperatures and enable utilization of piezomagnetic
QDs for high-density magnetic memory.

This work is supported by the US ONR, NSF-ECCS CAREER, the CCR
at SUNY Buffalo, and the Center for Nanophase Materials
Sciences, sponsored at ORNL by the Division of Scientific User
Facilities US DOE. We thank P. Hawrylak and D.~G. Austing
for stimulating discussions.


\begin{thebibliography}{99}
\bibitem{complex}
The size of the Hamiltonian matrix to be diagonalized for just
10 Mn (spin-5/2) is $6^{10} \approx 6 \times 10^7$.
\end{thebibliography}
\end{document}